\documentclass[twocolumn,superscriptaddress,showpacs,preprintnumbers,amsmath,amssymb,floatfix]{revtex4}
\usepackage{hyperref}
\usepackage{graphicx}
\usepackage{epsfig}

\newcommand{\be}{\begin{equation}}
\newcommand{\ee}{\end{equation}}
\newcommand{\ba}{\begin{array}{rcl}}
\newcommand{\ea}{\end{array}}
\newcommand{\ds}{\displaystyle}
\newcommand{\md}[1]{\left|#1\right|}

\begin{document}

\title{A Markov State Modeling analysis of sliding dynamics of a 2D model }

\author{M. Teruzzi}
\affiliation{SISSA, Via Bonomea 265, I-34136 Trieste, Italy}
\author{F. Pellegrini}
\affiliation{SISSA, Via Bonomea 265, I-34136 Trieste, Italy}
\affiliation{CNR-IOM Democritos National Simulation Center, Via Bonomea 265, I-34136
Trieste, Italy}
\author{A. Laio}
\affiliation{SISSA, Via Bonomea 265, I-34136 Trieste, Italy}
\affiliation{International Centre for Theoretical Physics (ICTP), Strada Costiera
11, I-34151 Trieste, Italy}
\author{E. Tosatti}
\affiliation{SISSA, Via Bonomea 265, I-34136 Trieste, Italy}
\affiliation{CNR-IOM Democritos National Simulation Center, Via Bonomea 265, I-34136
Trieste, Italy}
\affiliation{International Centre for Theoretical Physics (ICTP), Strada Costiera
11, I-34151 Trieste, Italy}

\date{\today}
\begin{abstract}
Non-equilibrium Markov State Modeling (MSM) has recently 
been proposed [Phys. Rev. E 94, 053001 (2016)] as a possible route to
construct a physical theory of sliding friction
from a long steady state atomistic simulation: the approach builds
a small set of collective variables, which obey a transition-matrix
based equation of motion, faithfully describing the slow motions
of the system.
A crucial question is whether this approach can be extended from
the original 1D small size demo to larger and more realistic size
systems, without an inordinate increase of the number and complexity
of the collective variables. Here we present a direct application
of the MSM scheme to the sliding of an island made of over 1000 harmonically
bound particles over a 2D periodic potential. Based on a totally unprejudiced
phase space metric and without requiring any special doctoring, we
find that here too the scheme allows extracting a very small number
of slow variables, necessary and sufficient to describe the dynamics
of island sliding.

\end{abstract}


\maketitle

\section{Introduction}
Sliding friction between
solid bodies, among the most basic and pervasive phenomena in physics
and in our everyday experience, can be measured, simulated but --
disappointingly -- not yet formulated theoretically. By that we mean
that even in the purely classical sliding of a body on another there
is no unprejudiced way of identifying and determining a handful of
variables (as opposed to the $\sim{10}^{23}$ atomic coordinates and
velocities) that obey a well defined equation of motion describing
the essence of the frictional process. The burgeoning area of 
nanofriction~\cite{vanossi2013}, where realistic simulations are often possible,
has made if anything this theoretical vacuum even more blatant. In
a recent pubblication we proposed~\cite{Pellegrini2016} that Markov
State modeling (MSM) -- a probabilistic approach commonly applied
to characterize the kinetics of systems characterized by an equilibrium
measure~\cite{Noe2009,Schwantes2014,Noe2013,Bowman2014,Schutte2015}
-- can be extended and used for the strongly non-equilibrium, non--linear
problem of sliding friction. The approach was demonstrated in a simple
1D toy model, a 10-atom Frenkel Kontorova model~\cite{Frenkel1938}
where, despite the difficulty represented by a time--growing phase
space, non-equilibrium MSM was shown to describe adequately the forced
dynamics of steady-state sliding friction. The probabilistic analysis
of a long steady-state frictional simulation and the choice of a metric
led to the recognition of Markovian evolution in phase space, to the
identification of a few slow collective variables (``excitations'')
describing the events occurring in the course of sliding, and to the
construction of a transfer--matrix--dictated model of the time evolution
of probabilities. This approach represents in our view a first step
towards a theory of friction, and a methodological advance of very
significant importance.

Here we showcase the first application of the MSM approach to a more
realistic frictional system. We choose for this purpose the sliding
of a two-dimensional (2D) island of more than $1000$ particles,
harmonically interacting at a spacing that is incommensurate with
respect to a periodic 2D substrate potential. We consider different
sliding regimes, including the ``superlubric'' and, at the opposite
limit, the pinned regime. The results are rewarding: firstly, and
most importantly, MSM again identifies an extremely small set of significant
variables, despite the totally generic choice of metric and the much
larger dimensionality of the phase space in which the original model
is defined. This handful of variables in turn describes without any
built-in prejudice the main slow time-dependent frictional events,
including superlubric soliton flow and atomic stick-slip frictional
sliding of the island, in the two extreme and opposite regimes of
weak and strong potential. 

\section{Markov State Modeling}
The Markov State Modeling procedure starts from a classical molecular dynamics
simulation of the sliding system, long enough to explore all relevant 
configurations in phase space a sufficiently large number of times. 
Structurally similar configurations are then grouped in a finite number of
{\em microstates}, which will serve as a basis, through a clustering
(such as k--means~\cite{Perez-Hernandez2013}) or geometric technique.
This partitioning requires a {\em metric} in phase space to
quantify the similarity between configurations: the quality of
the partition will generally depend on this choice, to be made
with utmost physical care. 
While in real world applications it is considered mandatory to 
define the metric in a relevant subset of the coordinates (for
example, the coordinates of the solute), we will show that, in the
specific system considered in this work, one can carry on the procedure
even using a ``blind'' metric, defined taking into account all the
cooordinates of the system.

In equilibrium settings, the transition probability matrix between 
pairs of microstates (in a time $\tau$) would be equivalent
to a hermitian matrix (on account
of detailed balance) and have a unique eigenvalue equal to one,
whose eigenvector represent the equilibrium distribution, and all other 
real eigenvalues smaller than one. The eigenvectors of eigenvalues 
closest to one are associated with slow modes of the system evolution, 
while the smaller eigenvalues correspond to fast motions, expected
to be increasingly irrelevant. A clear gap between high and low
eigenvalues leads to a natural dimensional reduction~\cite{Deuflhard2004,
Weber2005}.

To study non-equilibrium problems such as friction, this procedure has 
been modified in several key points: the frictional dynamics does not 
reach an equilibrium, but instead reaches a steady state where the
configuration space grows (approximately) linearly with the simulation 
time, making sampling and clustering problematic. The solution we 
proposed is dividing the evolution in intervals, still long enough to 
be deemed equivalent between them, so that results from each interval 
can be cumulated on top of one another. Stability of the results against 
extension of the time interval indicates the validity of the procedure.
Care should be taken in dealing with the transition probability matrix 
from this steady--state evolution under forcing, which is 
non-hermitian~\cite{Pellegrini2016}. 
Moreover, since the phase space metric contains a large number
of microscopic variables, we do not build microstates by the tessellation
techniques used in standard MSM~\cite{Prinz2011}, but use instead
a recently proposed clustering algorithm~\cite{Rodriguez2014}, which
associates a microstate to each meaningful peak of the probability
distribution in the coordinate space associated with the metric.

The main goal of this contribution is demonstrating that this procedure
works for a much more realistic model than the one considered in 
ref.~\cite{Pellegrini2016}: the sliding of a 2D Frenkel Kontorova island
including approximately 1000 atoms on a periodic incommensurate potential.

\section{The 2D Frenkel-Kontorova model}

\begin{figure}
\centering 
\includegraphics[width=0.5\textwidth]{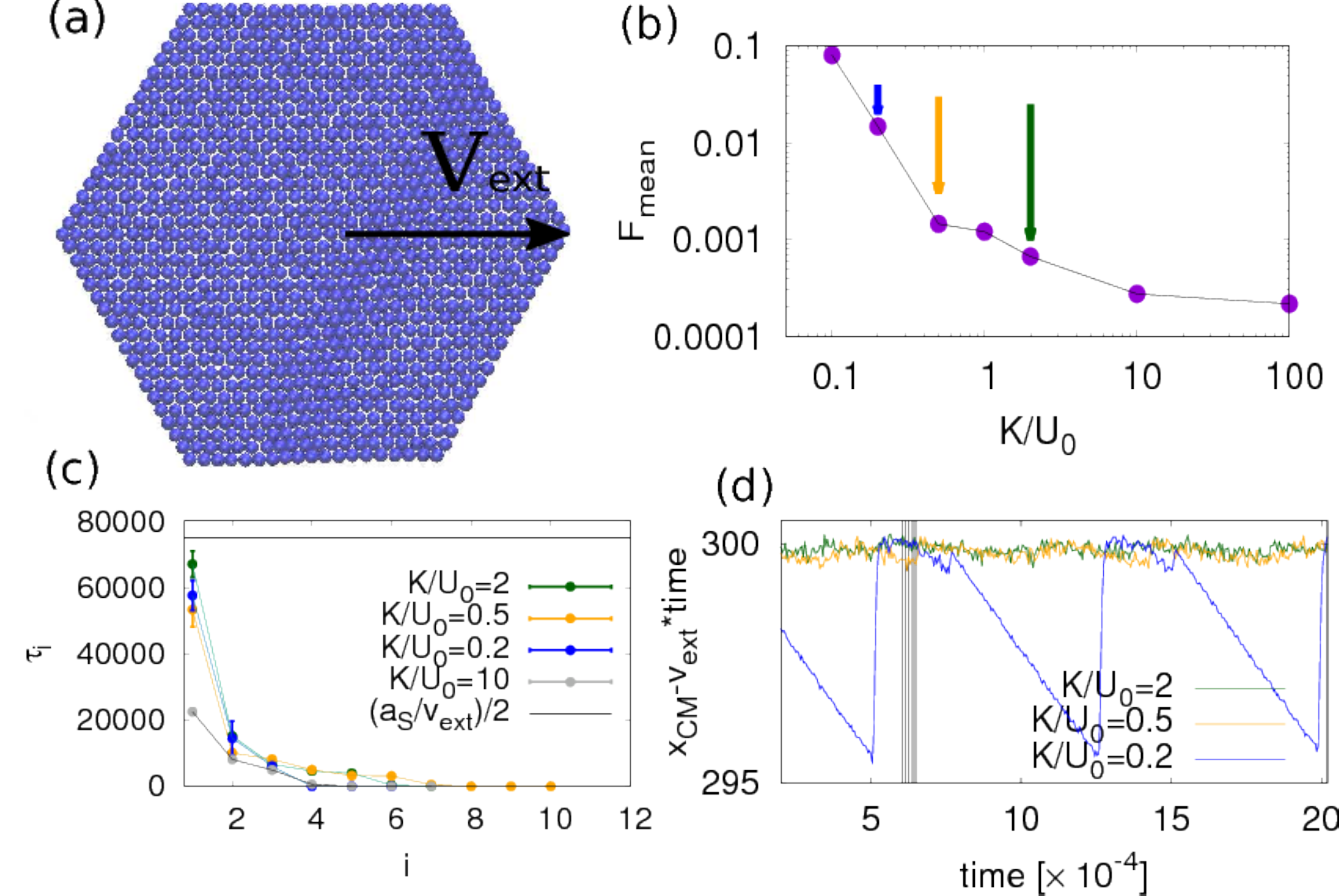} 
\caption{\label{Fig1}
(a) Schematic of the Frenkel-Kontorova island sliding on a generic 
2D incommensurate triangular potential.
(b) Average friction force in the steady state regime for simulations
with varying ratio $K/U_0$, highlighting the transition from free
sliding to stick--slip. The arrows indicate the sample parameters chosen
to compare the results of our method, with a color code we will keep 
throughout all figures. 
(c) Slowest timescales after transition matrix diagonalization for the
different regimes. The full line represents half of the time required
for the external force to move the particles to successive 
potential minima.
(d) Sample evolution for the different regimes: we show the deviations 
of the center of mass positions from the free sliding corresponding to 
an infinitely stiff island. The vertical dashed lines are spaced like the 
time lag $\tau=1100$ chosen to build the MSM to represent sampled configurations.}
\end{figure}

Our present study focuses on a two-dimensional FK model~\cite{Braun1998},
Fig.~\ref{Fig1}(a). We consider a hexagonal island of $N=1027$
classical particles, internally arranged as a triangular lattice,
dragged by a force applied on the center of mass, which causes it
to slide over a triangular potential $V(x,y)=U_{0}\left[2\cos\left(\frac{2\pi x}{a_{S}}\right)
\cos\left(\frac{2\pi y}{\sqrt{3}a_{S}}\right)+\cos\left(\frac{4\pi y}{\sqrt{3}a_{S}}\right)\right]$.
Nearest neighbor harmonic springs of stiffness $K$ 
link the particles of mass $m$ and positions $\mathbf{r}_{i}=(x_{i},y_{i})$
whose equilibrium spacing $a_{H}$ is incommensurate with the periodic
potential: $a_{S}/a_{H}\sim1.07$. Each particle is dragged by a spring
of constant $\kappa$ moving with constant velocity $v_{{\rm ext}}$.
Particle motion obeys an overdamped Langevin dynamics (large damping
$\gamma$), in a bath of inverse temperature $\beta=1/k_{B}T$:
\be\ba
\mathbf{r}_l^{t+dt}&=&\ds \mathbf{r}_l^t+
\left[\frac{1}{\gamma m}\nabla V(\mathbf{r}_l^t)
+\frac{\kappa}{\gamma m}\left(v_{\rm ext}t-\frac{1}{N}\underset{j}{\sum}x_j^t\right)+\right.\\
&&\ds\left.-\frac{K}{\gamma m}\sum_{j\in\mathrm{NN}}
(\mathbf{r}_l^t-\mathbf{r}_{j}^t)
\right] dt
+\sqrt\frac{2 dt}{\gamma m \beta}\mathbf{f}^t,
\ea\ee
where $\mathbf{f}^{t}$
is an uncorrelated Gaussian distribution and $dt$ is the elementary
time step (here $dt=0.1$, $m=1$, $\gamma=1$, $\beta=100$, $\kappa=0.01$, $v_{\rm ext}=0.0001$).

In a temperature and parameter regime where the island does not rotate,
its sliding mechanics retains some similarity to 1D sliding~\cite{Braun1998}.
In the weak potential limit the bulk of the island, characterized
by weak solitons (small deviations of the interparticle distance from
the equilibrium value) which form a moir\'e pattern over the incommensurate
potential, is structurally lubric (``superlubric''). Upon sliding
in this regime the solitons flow unhindered, and the only source of
pinning and static friction is actually provided by the island edge~\cite{varini2016}.
In the opposite strong potential limit the solitons, no longer weak,
are strongly entrenched, and the whole island is pinned, with a bulk
static friction independent of edges. Under the external spring--transmitted
force, the island sliding in this regime will alternate long ``sticking''
periods during which particles are close to their respective potential
minima, to fast slips during which one or more lattice spacings are
gained. This kind of atomic stick--slip motion is well established
for, e.g., the sliding of an Atomic Force Microscope tip on a crystal
surface~\cite{vanossi2013} -- of course involving in that case three-dimensional
displacements of larger complexity. A slip event always involves the
flow of either pre--existing solitons or of newly created ones that
enable the system to slide faster.

Our input for the MSM procedure is a long steady-state trajectory of
the island motion, obtained by integrating these equations for $\sim{10}^{8}$
timesteps for a slow external velocity, but far from the linear
response regime. 
Values of $K/U_{0}$ were chosen so as to straddle between
and beyond the weak ($K/U_{0}\geq2$) and strong ($K/U_{0}\leq0.2$)
potential regimes.

The average friction force $F_{\rm mean} = \kappa \left\langle v_{ext} t 
- \frac{1}{N}\sum_l x_{l}\right\rangle$
obtained from the simulation as a function of the ratio $K/U_{0}$
can be seen in Fig.~\ref{Fig1}(b), where the crossover
from a superlubric to a pinned regime is clearly reflected. We focus
our study on three different values of the parameters representative
of these different regimes, as can be hinted by looking at the evolution
of the position of the center of mass shown in Fig.~\ref{Fig1}(d).

\begin{figure}
\centering 
\includegraphics[width=0.5\textwidth]{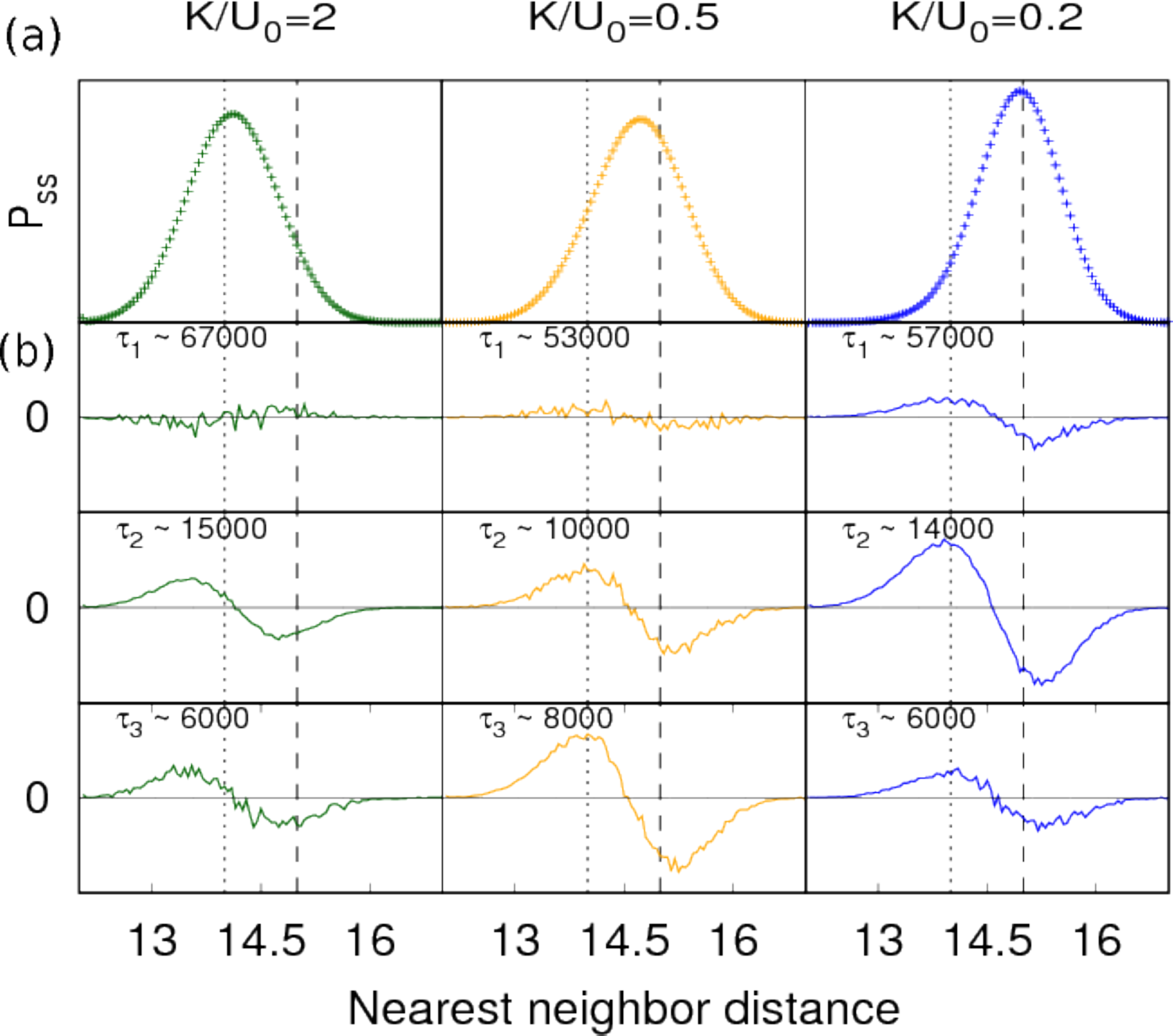} 
\caption{\label{Fig2}
(a) Steady state probability distribution of the nearest neighbor 
distances for the three regimes. Vertical dotted lines represent
the rest interparticle (``harmonic'') distance, while the dashed ones the substrate lattice spacing.
(b) The perturbations $g_i$ (see Eq.~\ref{EqForgi}) estimated for the first three excited states. The observable $O$ in Eq.~\ref{EqForgi} is here 
the nearest neighbor distance. 
To compute $P(O | \alpha)$ 100 intervals have been chosen.
}
\end{figure}

\section{Implementing the frictional MSM}

The protocol begins by defining a metric, measuring distances between
configurations in phase space. Since we want to remain as unprejudiced
as possible, we adopt the simplest, most generic and bias--free metric,
and define the distance between two configurations $s$ and $t$ as:
\be
d_{st}=\left[({\mathbf{r}}^{s}_{\rm CM}-{\mathbf{r}}^{t}_{\rm CM})_{\text{mod }2}\right]^2
+\left[\overset{N}{\underset{l=1}{\sum}}({\mathbf{r}}^{s}_{l}
-{\mathbf{r}}^{t}_{l})_{\text{mod }2}\right]^2. \label{Eq:dist_dB} 
\ee

The microstates were built using the Density Peak algorithm~\cite{Rodriguez2014}. This approach
requires only defining a distance between the configurations, here estimated using Eq.~\ref{Eq:dist_dB}.
Based on this definition, the approach automatically finds the peaks in the probability distribution
in the space of the coordinates in which the distance is defined. Here, following our previous work~\cite{Pellegrini2016}, we     
identify the microstates used for building the MSM with the density peaks.

We used samples of $N_{\rm conf}\sim{10}^{4}$ configurations (separated
by the lagtime $\tau$) and clustered them using the metric~(\ref{Eq:dist_dB}).
The order of magnitude of the lagtime $\tau$ has been chosen in order
to describe the stick (and slip) events of the system. The optimal
lagtime $\tau=1100$ was determined after some convergence checks
resembling those carried out in the previous work~\cite{Pellegrini2016}:
In particular, we verified that the relevant 
timescales stay within the statistical error in Fig.~\ref{Fig1} when doubling or 
halving the lagtime.  We also verified that by using the Core Set MSM 
approach~\cite{Schutte2011} the influence of the time lag on the timescales is further reduced. 
This indicates  that we are indeed far from the non-Markovian regime.

Given these $\{c_{\alpha},\alpha=1,\dots,n_{c}\}$ microstates, we
can construct a discretized, coarse--grained Transfer
Operator (TO)~\cite{Bowman2014}:
if $\Pi^{\tau}(X\rightarrow X')$ is the probability to go from a 
configuration $X$ at time $t$ to $X'$ at time $t+\tau$,
a finite $n_{c}\times n_{c}$ Transfer Matrix (TM) can be built 
by estimating the probability to go from $c_{\alpha}$ to $c_{\beta}$ 
in time $\tau$: $\Pi_{\alpha\beta}^{\tau}=\int_{X\in c_{\alpha}}
\int_{X'\in c_{\beta}}{\rm d}X{\rm d}X'P(X)\Pi^{\tau}(X\rightarrow X')$.
This TM contains less detail than $\Pi^{\tau}$, but 
it can be sampled in finite time. In principle, $\Pi_{\alpha\beta}^{\tau}$ 
depends on the choice of $\tau$, but an optimal value for this parameter
can be chosen.
We call $\{\lambda_{i}\}$ the eigenvalues of the TM and $\{\vec{\chi}_{i}\}$ 
its left eigenvectors. Since detailed balance does not hold, the TM is 
not symmetric and the eigenvalues can be complex, however $\md{\lambda_{i}}\le1$ 
is still guaranteed; the eigenvalue of largest modulus
is still $1$ and unique if the evolution is ergodic. 
The eigenvector $\vec{\chi}_{0}$ represents the
steady state distribution, while the eigenvectors $\vec{\chi}_{i}$ with 
$\md{\lambda_{i}}\simeq1$ form the so--called Perron Cluster~\cite{Deuflhard2004}. 
They characterize the long--lived perturbations to the steady state, decaying with
characteristic times $\tau_{i}=-\tau/\ln\md{\lambda_{i}}\gg\tau$,
while oscillating with period $\tau/\arctan({\rm Im}\lambda_{i}/{\rm Re}\lambda_{i})$.

To better characterize the eigenmodes $\chi_{i}$ it is useful to consider
a system prepared in the mixed state $P_{\alpha}^{0}$ (probability vector 
to be in $c_{\alpha}$) at $t=0$ and the evolution of the probability 
distribution $P\left(O,t\right)$ of an observable $O$ as a function of time.
We have: 
\be\label{eq:Pexp}
P\left(O,t\right)=P^{\rm ss}\left(O\right)+\sum_{i>0}f_{i}g_{i}\left(O\right)e^{-t/\tau_{i}},
\ee
where $f_i=\sum_\alpha\chi_i^\alpha P^0_\alpha/P^{\rm ss}_\alpha$ 
accounts for the initial condition and 
\be
g_{i}\left(O\right)=\sum_{\alpha}\chi_i^\alpha P(O | \alpha),
\label{EqForgi}
\ee
where $P(O|\alpha)$ is the
probability distribution of $O$ in microstate $\alpha$, $P^{{\rm ss}}(O)=g_{0}(O)$
the steady state distribution of $O$, and $P_{\alpha}^{{\rm ss}}$
the steady state probability to visit microstate $\alpha$. The $g_{i}(O)$
for $i>1$ represent ``perturbations'' of $P^{{\rm ss}}(O)$, each
decaying within the lifetime $\tau_{i}$.
While the expansion~\eqref{eq:Pexp} is meaningful only for a given
starting configuration, the analysis of the shape of these functions
(regardless of their amplitude) provides a direct insight into the 
nature of the slow eigenmodes.
One can therefore turn back to observables deemed relevant in the 
original system and estimate their influence in the relevant 
dynamical modes of the system in order to gain insight on their
characteristics.

\begin{figure}
\centering 
\includegraphics[width=0.5\textwidth]{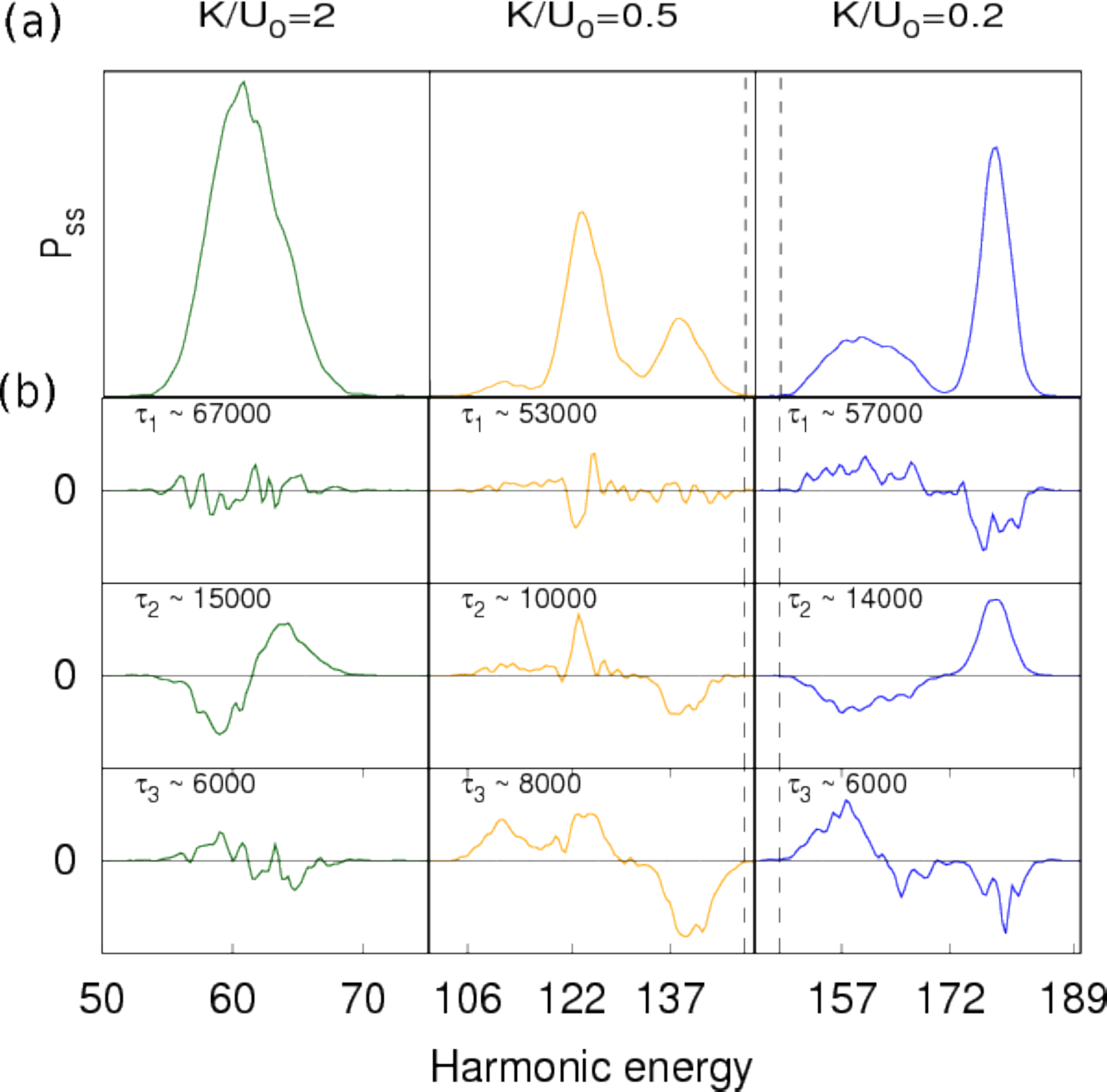} 
\caption{\label{Fig3}
(a) Steady state probability distribution of the harmonic energy 
for the three regimes. The vertical dashed lines represent
the harmonic energy for an island completely relaxed to the substrate 
lattice spacing. In this case we applied a running average to the data, 
both for the steady--state distribution and the corrections.
The switch from a single peak of the hard island to a multiplicity of 
peaks for the medium to soft island is a direct evidence of the more 
elaborate sliding dynamics of the latter.
(b) The perturbations $g_i$ (see Eq.~\ref{EqForgi}) estimated for the 
first three excited states. The observable $O$ in Eq.~\ref{EqForgi} 
is here the harmonic energy.
To compute $P(O | \alpha)$ 100 intervals have been chosen.
}
\end{figure}

\section{Observables}

We apply the described procedure to three evolutions of our model characterized
by different parameters $K/U_0$ as indicated in Fig.~\ref{Fig1}(b).
The time corresponding to the first largest eigenvalue
(computed from the real part of the eigenvalues, 
besides the eigenvalue $0$ associated with the steady--state,
while the imaginary part is much smaller and has been ignored)
can be seen in Fig.~\ref{Fig1}(c): for all cases we find
that the first implied timescale is approximately equal to 
$6\times{10}^{4}$, corresponding to roughly half the 
time $a_S/v_{\rm ext}$ required on average to move by
one lattice spacing.
This is consistent with the interpretation of the slowest mode as being 
related to the movement from one local minimum of the substrate to the next. We notice that
in the more extreme case $K/U_0=10$ the first relaxation time is  
faster, since the island is stiff  the  substrate does not play a role.
The successive timescales are almost an order of magnitude faster, 
and further insight is required for their interpretation.
We will presently analyze the $g$ functions of some relevant
physical observables of this frictional
system, in order to characterize these rapidly decaying states.

\subsection{Nearest neighbour distance}

As a first observable we consider the nearest neighbor distance between 
all particle pairs. The steady--state distributions (see Fig.~\ref{Fig2}(a))
show the expected trend: while the $K/U_0=2$,case has a peak centered on the 
harmonic equilibrium distance 
reflecting the hard island's nature during structurally lubric sliding~\cite{vanossi2013},
the opposite case $K/U_0=0.2$ is centered on a distance commensurate with the 
substrate, reflecting the soft island's strong adhesion to the external potential. 
For $K/U_0=0.5$ the situation is intermediate.
The excited states complete this description (see Fig.~\ref{Fig2}(b)): 
the first excited states show little correction to the 
steady--state distribution, as the change of a whole lattice spacing has 
only a minor influence on the nearest neighbor distribution, while the second 
and third excited states display a significant change. Indeed,
the latter correspond to internal relaxations of the island not 
associated with the collective sliding.
In the specific case these corrections  are related to the 
formation/destruction of incommensurate solitons
induced on the island by the external potential~\cite{varini2016}.

This observable lacks the ability to clearly distinguish
between the excited states. We therefore considered a
more extensive observable, able to highlight more  differences.

\begin{figure}
\centering 
\includegraphics[width=0.5\textwidth]{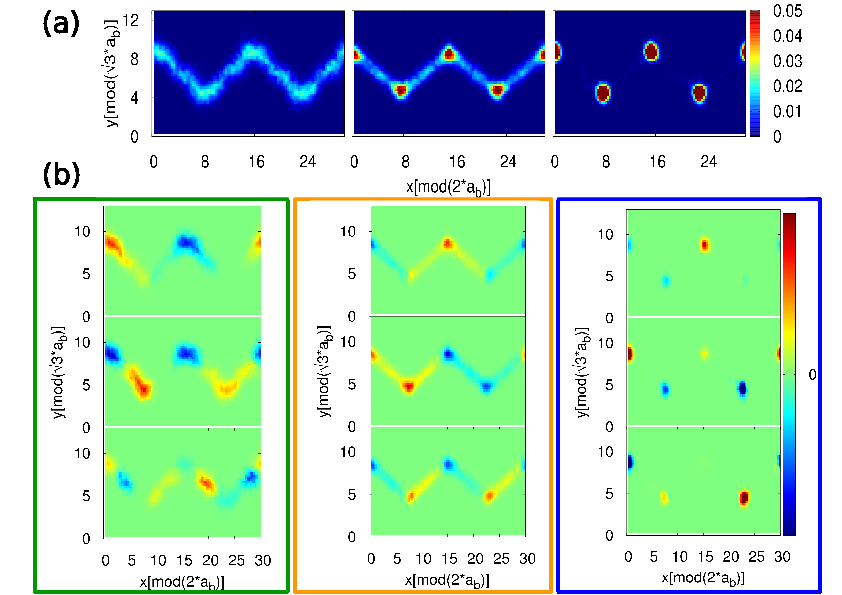} 
\caption{\label{Fig4}
(a) Steady state 2D probability distribution of all single particle positions 
(excluding edges) for the three regimes. Position along x and y is taken modulus 
1 and 2 lattice spacings, respectively.  
The zig-zag pattern reflects the motion of the whole which, while pulled along $X$, 
moves between neighboring potential minima that are $\pm 60^\circ$ off. 
(b) The perturbations $g_i$ (see Eq.~\ref{EqForgi}) estimated for the first 
three excited states. The observable $O$ in Eq.~\ref{EqForgi} is here 
the single particle positions.
Note the evolution for decreasing $K/U_0$, from smooth flow to sharp hopping betweem minima.
To compute $P(O | \alpha)$ 100 intervals have been chosen for x and 50 for y.
}
\end{figure}

\subsection{Harmonic energy}

We now consider the distribution of the total harmonic energy of the island 
$\frac{K}{2}\langle\sum_{\langle i,j\rangle\in\mathrm{NN}}(\mathbf{r}_i^t-\mathbf{r}_{j}^t)^2 \rangle$.
Fig.~\ref{Fig3}(a) shows the steady--state distributions, 
clearly highlighting the richer information encoded by this observable.
In the stiff $K/U_0=2$ case the distribution of this observable shows a single peak, while in the softer
cases it acquires a more complex structure, related to the presence of a different number of
solitons in the system. The corrections in Fig.~\ref{Fig3}(b) highlight how the different modes
(besides the first one, as previously noted) are related to different relative weights in these
soliton distributions, representative of the different dynamics of each regime:
while in the stiff case the few defects merely slide through the island during the motion, 
leaving their population unchanged, for the softer islands the stick-slip motion is achieved through
the creation of new solitons at the edges and their relatively fast propagation, leading to a complex
time dependence of their population.

\subsection{Single particle positions}

To gain additional insight in the nature of the slow dynamical modes we
now consider the probability distribution of the position of a single 
particle (not on the border)
as a function of its position $x$ and $y$, modulus the periodicity of the substrate. 

The steady state positional distribution of (Fig.~\ref{Fig4}(a)) again  shows 
how the increase of $U_0$ leads from a smooth distribution over the continuous 
transition path from a minimum to the next, to an increasingly 
peaked distribution in the potential minima. Therefore if for $K/U_0=2$ 
the particle performs a rather smooth 
zig-zag path between successive potential minima,
In the intermediate case ($K/U_0=0.5$) these positions
are much more probable, eventually becoming dominant
for $K/U_0=0.2$ where the distribution reduces essentially to sharp peaks. 
The excited state effect on the particle position distribution is
shown in Fig.~\ref{Fig4}(b). 
While the first excitation is clearly related to the single period shift,
as mentioned earlier, the second excited state 
shows that the particle jumps among successive minima in the zig-zag path.

This shorter periodicity was not visible in the previous
observables as it is not shared by all particles. The third excited state, 
finally, reflects the particle position  
probability perturbation caused by the ``slip'' events, which are 
characteristic and strong for the softer island, as in the previous analysis.

\subsection{Work distribution}

\begin{figure}
\centering 
\includegraphics[width=0.5\textwidth]{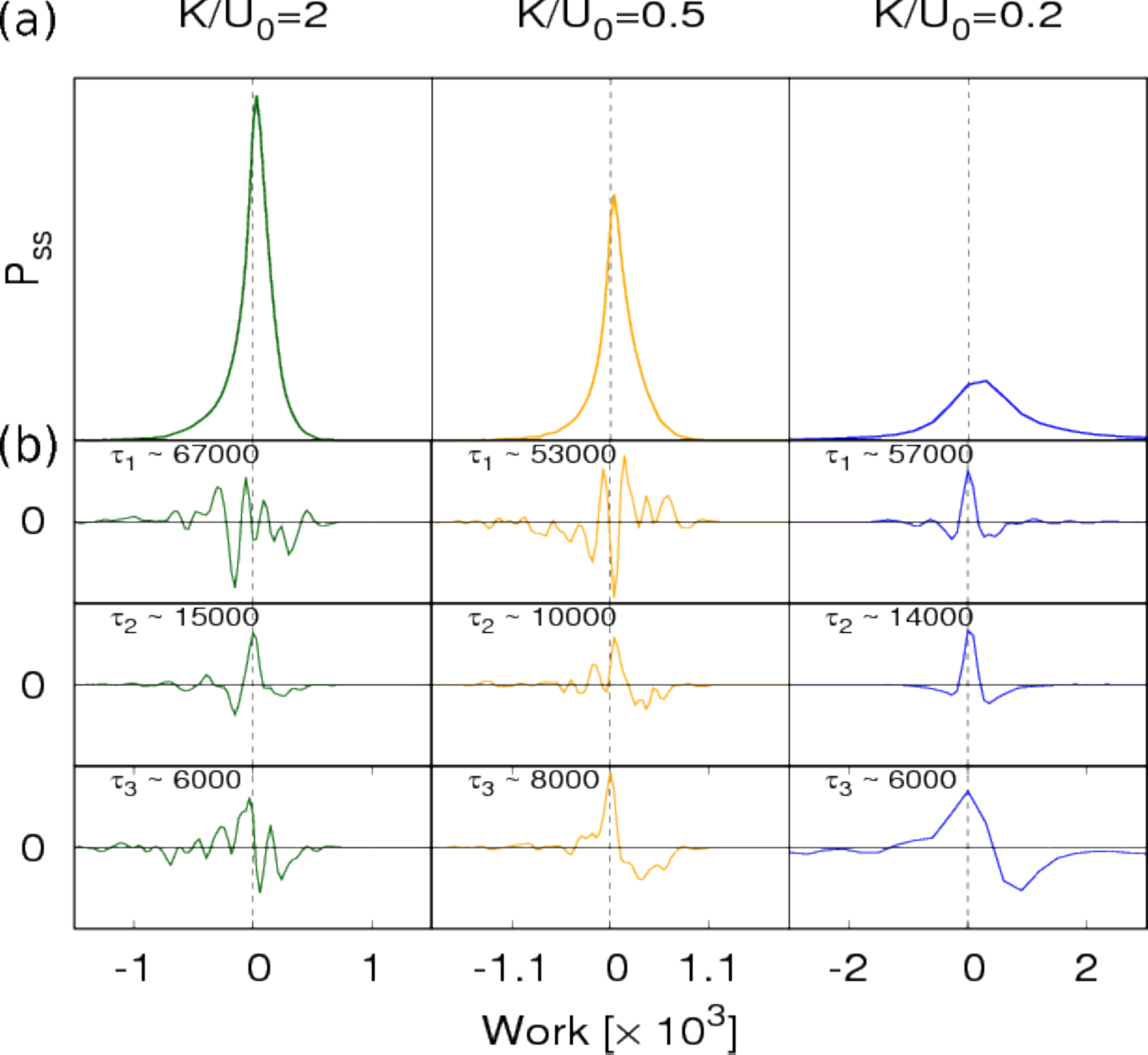} 
\caption{\label{Fig5}
(a) Steady state probability distribution of work 
for the three regimes. In this case we applied a running average to the data, both 
for the steady--state distribution and the corrections.
(b) The perturbations $g_i$ (see Eq.~\ref{EqForgi}) estimated for the first three 
excited states. The observable $O$ in Eq.~\ref{EqForgi} is here the work.
To compute $P(O | \alpha)$ 100 intervals have been chosen.
}
\end{figure}

As a final observable, relevant to the description of a frictional model, we consider 
the instantaneous work done on the system by the external force in a single
timestep $\tau$: 
$W_t = \kappa\sum_l (v_{\rm ext}t-x_l^t) (x_l^{t+\tau}-x_l^t)$,
shown in Fig.~\ref{Fig5}(a).
(Notice that this quantity depend on the successive position at timestep
$t$ and $t+\tau$).
The steady-state work distribution  $P_{ss}(W)$ is centered on $ \langle W \rangle $, 
a value evolving from near zero to larger values as one goes from $K/U_0=2$ to $K/U_0=0.2$.
At the same time $P_{ss}(W)$  develops an increasing asymmetry 
with a broader and broader tail around positive values of work.
Both features are related to the increase of dissipation as the substrate
corrugation increases.
As for the previous observables, this steady-state level information is straight 
from the simulation and does not need the MSM analysis. 
Now however we can examine what the excitations do. 

As seen in  Fig.~\ref{Fig5}(b) for $K/U_0=2$ the excitations show just noise, which tells 
us that the slider moves as a whole, as characteristic of the superlubric sliding in this regime.
The notable exception is however the second excitation, showing a forward jump. This marginal 
stick-slip behaviour is actually due to the weak but nonzero pinning caused by the island edge 
that hinder the entrance and exit of solitons~\cite{varini2016}, a subtle but real feature 
which in this case is efficiently and unbiasedly uncovered through this excitation.
  
As we move towards smaller and smaller $K/U_0$ and the island softens, all excitations 
gradually come into play.
In the final stick--slip regime, modifications in the 
soliton structure are strongly related to an increase in the positive tail of the work
distribution, highlighting the mechanism behind the increased friction coefficient.

\section{Conclusions}

The Markov State Model method, so far developed for the equilibrium
evolution of large--scale molecular systems, can be naturally extended
to non-equilibrium dynamics under the action of external forces. Among
non--equilibrium phenomena, the physics of sliding friction is in
bad need of a description, with coarse--grained variables and their
time evolution constructed in the least prejudiced manner. We have
shown here that application of this technique to a realistic model
involving a mesoscopically large sliding system is possible and fruitful.

Three important conclusions that were not a priori granted deserve
being underlined. The first is that no particularly clever or savvy
choice of the metric is necessary: the very naive choice of considering
the distance between all the particles of the sliding island works
very well. Since the metric is so simple, the kinetic model that is
obtained is likely to be accurate. The second, and equally remarkable
result is that despite many thousands of atomistic degrees of freedom,
the procedure allows selecting just very few slow variables, automatically
eliminating all other fast irrelevant variables. The third is that
the slow variables, once examined at the end, are found to make a
lot of sense when confronted with the actual frictional physics of
the system, be it superlubric or stick--slip. These gratifying bottomlines
provide a strong encouragement towards the future use of the MSM for
the theoretical description of sliding friction.


\section*{ACKNOWLEDGMENTS}

Work carried out under ERC Advanced Research Grant N.~320796 -- MODPHYSFRICT. COST Action
MP1303 is also acknowledged. Early discussions with Fran\c{c}ois Landes are gratefully acknowledged.


\begin{thebibliography}{25}
\expandafter\ifx\csname natexlab\endcsname\relax\def\natexlab#1{#1}\fi
\expandafter\ifx\csname bibnamefont\endcsname\relax
  \def\bibnamefont#1{#1}\fi
\expandafter\ifx\csname bibfnamefont\endcsname\relax
  \def\bibfnamefont#1{#1}\fi
\expandafter\ifx\csname citenamefont\endcsname\relax
  \def\citenamefont#1{#1}\fi
\expandafter\ifx\csname url\endcsname\relax
  \def\url#1{\texttt{#1}}\fi
\expandafter\ifx\csname urlprefix\endcsname\relax\def\urlprefix{URL }\fi
\providecommand{\bibinfo}[2]{#2}
\providecommand{\eprint}[2][]{\url{#2}}

\bibitem[{\citenamefont{Vanossi et~al.}(2013)\citenamefont{Vanossi, Manini,
  Urbakh, Zapperi, and Tosatti}}]{vanossi2013}
\bibinfo{author}{\bibfnamefont{A.}~\bibnamefont{Vanossi}},
  \bibinfo{author}{\bibfnamefont{N.}~\bibnamefont{Manini}},
  \bibinfo{author}{\bibfnamefont{M.}~\bibnamefont{Urbakh}},
  \bibinfo{author}{\bibfnamefont{S.}~\bibnamefont{Zapperi}}, \bibnamefont{and}
  \bibinfo{author}{\bibfnamefont{E.}~\bibnamefont{Tosatti}},
  \bibinfo{journal}{Reviews of Modern Physics} \textbf{\bibinfo{volume}{85}},
  \bibinfo{pages}{529} (\bibinfo{year}{2013}).

\bibitem[{\citenamefont{Pellegrini et~al.}(2016)\citenamefont{F. Pellegrini, 
F.P. Landes, A. Laio, S. Prestipino, and E. Tosatti}}]{Pellegrini2016}
\bibinfo{author}{\bibfnamefont{F.}~\bibnamefont{Pellegrini}},
  \bibinfo{author}{\bibfnamefont{F.P.}~\bibnamefont{Landes}},
  \bibinfo{author}{\bibfnamefont{A.}~\bibnamefont{Laio}},
  \bibinfo{author}{\bibfnamefont{S.}~\bibnamefont{Prestipino}}, \bibnamefont{and}
  \bibinfo{author}{\bibfnamefont{E.} \bibnamefont{Tosatti}},
  \bibinfo{journal}{Physical Review E} \textbf{\bibinfo{volume}{94}},
  \bibinfo{pages}{053001} (\bibinfo{year}{2016}).
  
\bibitem[{\citenamefont{No{\'{e}} et~al.}(2009)\citenamefont{No{\'{e}},
  Sch{\"{u}}tte, Vanden-Eijnden, Reich, and Weikl}}]{Noe2009}
\bibinfo{author}{\bibfnamefont{F.}~\bibnamefont{No{\'{e}}}},
  \bibinfo{author}{\bibfnamefont{C.}~\bibnamefont{Sch{\"{u}}tte}},
  \bibinfo{author}{\bibfnamefont{E.}~\bibnamefont{Vanden-Eijnden}},
  \bibinfo{author}{\bibfnamefont{L.}~\bibnamefont{Reich}}, \bibnamefont{and}
  \bibinfo{author}{\bibfnamefont{T.~R.} \bibnamefont{Weikl}},
  \bibinfo{journal}{Proceedings of the National Academy of Sciences of the
  United States of America} \textbf{\bibinfo{volume}{106}},
  \bibinfo{pages}{19011} (\bibinfo{year}{2009}).

\bibitem[{\citenamefont{Schwantes et~al.}(2014)\citenamefont{Schwantes,
  McGibbon, and Pande}}]{Schwantes2014}
\bibinfo{author}{\bibfnamefont{C.~R.} \bibnamefont{Schwantes}},
  \bibinfo{author}{\bibfnamefont{R.~T.} \bibnamefont{McGibbon}},
  \bibnamefont{and} \bibinfo{author}{\bibfnamefont{V.~S.} \bibnamefont{Pande}},
  \bibinfo{journal}{The Journal of Chemical Physics}
  \textbf{\bibinfo{volume}{141}}, \bibinfo{pages}{090901}
  (\bibinfo{year}{2014}).

\bibitem[{\citenamefont{No{\'{e}} and N{\"{u}}ske}(2013)}]{Noe2013}
\bibinfo{author}{\bibfnamefont{F.}~\bibnamefont{No{\'{e}}}} \bibnamefont{and}
  \bibinfo{author}{\bibfnamefont{F.}~\bibnamefont{N{\"{u}}ske}},
  \bibinfo{journal}{Multiscale Modeling {\&} Simulation}
  \textbf{\bibinfo{volume}{11}}, \bibinfo{pages}{635} (\bibinfo{year}{2013}).

\bibitem[{\citenamefont{Bowman et~al.}(2014)\citenamefont{Bowman, Pande, and
  No{\'{e}}}}]{Bowman2014}
\bibinfo{author}{\bibfnamefont{G.~R.} \bibnamefont{Bowman}},
  \bibinfo{author}{\bibfnamefont{V.~S.} \bibnamefont{Pande}}, \bibnamefont{and}
  \bibinfo{author}{\bibfnamefont{F.}~\bibnamefont{No{\'{e}}}}, in
  \emph{\bibinfo{booktitle}{An Introduction to Markov State Models and Their
  Application to Long Timescale Molecular Simulation}},
  \bibinfo{publisher}{Springer} (\bibinfo{year}{2014}).

\bibitem[{\citenamefont{Sch{\"{u}}tte and Sarich}(2015)}]{Schutte2015}
\bibinfo{author}{\bibfnamefont{C.}~\bibnamefont{Sch{\"{u}}tte}}
  \bibnamefont{and} \bibinfo{author}{\bibfnamefont{M.}~\bibnamefont{Sarich}},
  \bibinfo{journal}{The European Physical Journal Special Topics}
  \textbf{\bibinfo{volume}{18}} (\bibinfo{year}{2015}).

\bibitem[{\citenamefont{Frenkel and {T.A. Kontorova}}(1938)}]{Frenkel1938}
\bibinfo{author}{\bibfnamefont{Y.~I.} \bibnamefont{Frenkel}} \bibnamefont{and}
  \bibinfo{author}{\bibnamefont{{T.~A. Kontorova}}}, \bibinfo{journal}{Phys. Z.
  Sowietunion} \textbf{\bibinfo{volume}{13}} (\bibinfo{year}{1938}).

\bibitem[{\citenamefont{P{\'{e}}rez-Hern{\'{a}}ndez
  et~al.}(2013)\citenamefont{P{\'{e}}rez-Hern{\'{a}}ndez, Paul, Giorgino, {De
  Fabritiis}, and No{\'{e}}}}]{Perez-Hernandez2013}
\bibinfo{author}{\bibfnamefont{G.}~\bibnamefont{P{\'{e}}rez-Hern{\'{a}}ndez}},
  \bibinfo{author}{\bibfnamefont{F.}~\bibnamefont{Paul}},
  \bibinfo{author}{\bibfnamefont{T.}~\bibnamefont{Giorgino}},
  \bibinfo{author}{\bibfnamefont{G.}~\bibnamefont{{De Fabritiis}}},
  \bibnamefont{and} \bibinfo{author}{\bibfnamefont{F.}~\bibnamefont{No{\'{e}}}},
  \bibinfo{journal}{The Journal of Chemical Physics}
  \textbf{\bibinfo{volume}{139}}, \bibinfo{pages}{015102}
  (\bibinfo{year}{2013}).

\bibitem[{\citenamefont{Deuflhard and Weber}(2005)}]{Deuflhard2004}
\bibinfo{author}{\bibfnamefont{P.}~\bibnamefont{Deuflhard}} \bibnamefont{and}
  \bibinfo{author}{\bibfnamefont{M.}~\bibnamefont{Weber}},
  \bibinfo{journal}{Linear Algebra and its Applications}
  \textbf{\bibinfo{volume}{398}}, \bibinfo{pages}{161} (\bibinfo{year}{2005}).

\bibitem[{\citenamefont{Weber and Kube}(2005)}]{Weber2005}
\bibinfo{author}{\bibfnamefont{M.}~\bibnamefont{Weber}} \bibnamefont{and}
  \bibinfo{author}{\bibfnamefont{S.}~\bibnamefont{Kube}},
  \bibinfo{journal}{Lecture Notes in Computer Science}
  \textbf{\bibinfo{volume}{3695}}, \bibinfo{pages}{57} (\bibinfo{year}{2005}).

\bibitem[{\citenamefont{Prinz et~al.}(2011)\citenamefont{Prinz, Wu, Sarich,
  Keller, Senne, Held, Chodera, Sch{\"{u}}tte, and No{\'{e}}}}]{Prinz2011}
\bibinfo{author}{\bibfnamefont{J.-H.} \bibnamefont{Prinz}},
  \bibinfo{author}{\bibfnamefont{H.}~\bibnamefont{Wu}},
  \bibinfo{author}{\bibfnamefont{M.}~\bibnamefont{Sarich}},
  \bibinfo{author}{\bibfnamefont{B.}~\bibnamefont{Keller}},
  \bibinfo{author}{\bibfnamefont{M.}~\bibnamefont{Senne}},
  \bibinfo{author}{\bibfnamefont{M.}~\bibnamefont{Held}},
  \bibinfo{author}{\bibfnamefont{J.~D.} \bibnamefont{Chodera}},
  \bibinfo{author}{\bibfnamefont{C.}~\bibnamefont{Sch{\"{u}}tte}},
  \bibnamefont{and} \bibinfo{author}{\bibfnamefont{F.}~\bibnamefont{No{\'{e}}}},
  \bibinfo{journal}{The Journal of Chemical Physics}
  \textbf{\bibinfo{volume}{134}}, \bibinfo{pages}{174105}
  (\bibinfo{year}{2011}).

\bibitem[{\citenamefont{Rodriguez and Laio}(2014)}]{Rodriguez2014}
\bibinfo{author}{\bibfnamefont{A.}~\bibnamefont{Rodriguez}} \bibnamefont{and}
  \bibinfo{author}{\bibfnamefont{A.}~\bibnamefont{Laio}},
  \bibinfo{journal}{Science} \textbf{\bibinfo{volume}{344}},
  \bibinfo{pages}{1492} (\bibinfo{year}{2014}).

\bibitem[{\citenamefont{Braun and Kivshar}(1998)}]{Braun1998}
\bibinfo{author}{\bibfnamefont{O.~M.} \bibnamefont{Braun}} \bibnamefont{and}
  \bibinfo{author}{\bibfnamefont{Y.~S.} \bibnamefont{Kivshar}},
  \bibinfo{journal}{Physics Reports} \textbf{\bibinfo{volume}{306}},
  \bibinfo{pages}{1} (\bibinfo{year}{1998}).

\bibitem[{\citenamefont{Varini et~al.}(2015)\citenamefont{Varini, Vanossi, Guerra,
 Mandelli, Capozza,and Tosatti,}}]{varini2016}
\bibinfo{author}{\bibfnamefont{N.} \bibnamefont{Varini}},
  \bibinfo{author}{\bibfnamefont{A.} \bibnamefont{Vanossi}},
  \bibinfo{author}{\bibfnamefont{R.} \bibnamefont{Guerra}},
  \bibinfo{author}{\bibfnamefont{D.} \bibnamefont{Mandelli}},
\bibinfo{author}{\bibfnamefont{R.} \bibnamefont{Capozza}}, \bibnamefont{and}
  \bibinfo{author}{\bibfnamefont{E.} \bibnamefont{Tosatti}},
  \bibinfo{journal}{Nanoscale}
  \textbf{\bibinfo{volume}{7}}, \bibinfo{pages}{2093} (\bibinfo{year}{2015}).

\bibitem {Schutte2011}
\bibinfo{author}{\bibfnamefont{C.} \bibnamefont{Sch{\"{u}}tte}},
  \bibinfo{author}{\bibfnamefont{F.} \bibnamefont{No{\'{e}}}},
  \bibinfo{author}{\bibfnamefont{J.} \bibnamefont{Lu}},
  \bibinfo{author}{\bibfnamefont{M.} \bibnamefont{Sarich}},
  \bibnamefont{and} \bibinfo{author}{\bibfnamefont{E.} \bibnamefont{Vanden-Eijnden}},
  \bibinfo{journal}{The Journal of Chemical Physics}
  \textbf{\bibinfo{volume}{134}}, \bibinfo{pages}{204105}
  (\bibinfo{year}{2011}).

\end{thebibliography}
\end{document}